\documentclass[twocolumn,showpacs,preprintnumbers,amsmath,
  amssymb,showkeys,floatfix,prl,dvips]{revtex4}

\usepackage{graphics}
\usepackage{graphicx}
\usepackage{dcolumn}
\usepackage{bm}
\usepackage{verbatim}

\begin{document}

\title{Turbulent Diamagnetism in Flowing Liquid Sodium}

\author{E. J. Spence}
\altaffiliation[Current address: ]{Institut f\"ur Geophysik, 
ETH Z\"urich, CH-8093 Z\"urich, Switzerland
}
\author{M. D. Nornberg}
\altaffiliation[Current address: ]{Princeton Plasma Physics 
Laboratory, Princeton University, P.O. Box 451, Princeton, 
New Jersey 08543
}
\author{C. M. Jacobson} 
\altaffiliation[Current address: ]{Princeton Plasma Physics 
Laboratory, Princeton University, P.O. Box 451, Princeton, 
New Jersey 08543
}
\author{C. A. Parada}
\author{N. Z. Taylor}
\author{R. D. Kendrick}
\author{C. B. Forest}
\email{cbforest@wisc.edu}
\affiliation{
  Department of Physics,  University of Wisconsin--Madison,
  1150 University Avenue, Madison, Wisconsin 53706
}

\date{\today}

\begin{abstract}

The nature of Ohm's law is examined in a turbulent flow of liquid
sodium.  A magnetic field is applied to the flowing sodium, and the
resulting magnetic field is measured.  The mean velocity field of the
sodium is also measured in an identical-scale water model of the
experiment.  These two fields are used to determine the terms in Ohm's
law, indicating the presence of currents driven by a turbulent
electromotive force.  These currents result in a diamagnetic effect,
generating magnetic field in opposition to the dominant fields of the
experiment.  The magnitude of the fluctuation-driven magnetic field is
comparable to that of the field induced by the sodium's mean flow.

\end{abstract}


\maketitle

The magnetic fields of stars and planets are generated by the motion
of an electrically conducting plasma or liquid metal in the star or
planet's interior \cite{Rudiger}.  The motion of the fluid across an
existing seed magnetic field generates a motional electromotive force
(EMF) that drives currents, leading to a self-generated magnetic
field.  While large scale flow may induce much of the magnetic field,
there is another means by which magnetic field may be generated: the
interaction of velocity and magnetic field fluctuations
\cite{Krause_and_Raedler}.  If the magnetic and velocity fields are
separated into their mean (taken to be the temporally averaged value)
and fluctuating components,
$\mathbf{B}=\left<\mathbf{B}\right>+\mathbf{\tilde{B}}$ and
$\mathbf{V}=\left<\mathbf{V}\right>+\mathbf{\tilde{V}}$, then the mean
electric current obeys an Ohm's law of the form
\begin{equation}
\left<\mathbf{J}\right>=\sigma\left(\left<\mathbf{E}\right> +
\left<\mathbf{V}\right>\times\left<\mathbf{B}\right> +
\big<\mathbf{\tilde{V}}\times\mathbf{\tilde{B}}\big>\right),
\label{eqn:eqn1}
\end{equation}
where $\left<\mathbf{J}\right>$ is the average current density,
$\sigma$ the conductivity of the fluid, and $\left<\mathbf{E}\right>$
the average electric field.  The electric field plays a passive role
in the scenario studied here, merely maintaining
$\nabla\cdot\left<\mathbf{J}\right>=0$, and as such will not be
considered further.  There are two significant source terms in Eq.~1:
$\left<\mathbf{V}\right>\times\left<\mathbf{B}\right>$ represents the
EMF associated with the mean part of the velocity and magnetic fields,
while $\big<\mathbf{\tilde{V}} \times
\mathbf{\tilde{B}}\big>$ represents the EMF generated by the
fluctuating part of the fields; velocity and magnetic field
fluctuations can interact coherently to generate mean currents.

The possibility of such a turbulent EMF has long been recognized
\cite{Steenbeck.ZNT.1966, Moffatt, Krause_and_Raedler}.  Much of its
study has focused on those currents generated by helical velocity
field fluctuations \cite{Parker.AJ.1955b, Krause.ZNT.1967}, and
several non-simply-connected liquid-metal experiments have been
constructed to mimic such helical flows \cite{Steenbeck.SPD.1968,
Stieglitz.PF.2001}.  However, gradients in the intensity of the
fluctuations can also generate currents.  These currents effectively
expel magnetic flux from regions of high turbulence to low, resulting
in a diamagnetic effect \cite{Vainshtein.ZPMTF.1971, Parker.APJ.1975b,
Meneguzzi.PRL.1981, Vainshtein.GAFD.1983, Krause_and_Raedler,
Cattaneo.APJ.1994, Blanchflower.MNRAS.1998, Tao.MNRAS.1998}.  Such
flux expulsion may explain \cite{Boldyrev.APJ.2006} the weak magnetic
field at the center of the galactic core \cite{LaRosa.APJ.2005}
relative to the core's external flux tubes
\cite{Yusef-Zadeh.APJ.2003}, as well as concentrations of large scale
toroidal magnetic field at the base of the stellar convection
zone~\cite{Arter.MNRAS.1982, Nordlund.APJ.1992, Tao.APJ.1998}.
However, since fluctuation-driven fields cannot yet be isolated in
astrophysical or geophysical settings, whether turbulent EMFs play a
significant role in the production of such magnetic fields remains an
open question.

\begin{figure*}
\includegraphics[height=3in]{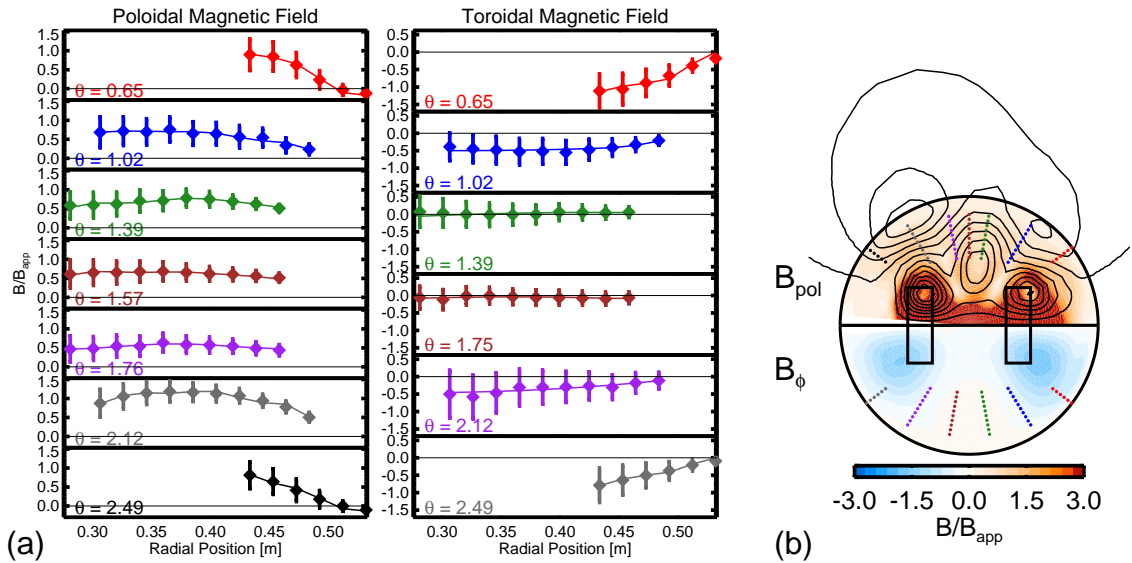}
\caption{Induced magnetic fields measured in the Madison Dynamo
Experiment.  (a) Mean measured poloidal and toroidal field values,
scaled to the magnitude of the applied field (50 Gauss), for an
impeller rotation rate of 1000 RPM.  The fit (solid lines) represents
values predicted by a spherical harmonic expansion fit to the data
(diamonds).  Error bars represent the RMS fluctuation levels of the
signals.  (b) The reconstructed field, with the axis of symmetry
oriented horizontally.  Streamlines of the poloidal field are in the
upper hemisphere and the toroidal field strength is in the lower
hemisphere.  Measurement positions are indicated with dots, and the
impeller positions are indicated with rectangles.}
\label{fig:mag_fit}
\end{figure*}

In this Letter we present the spatial structure of such
fluctuation-driven currents, as measured in the Madison Dynamo
Experiment.  The net result is a strong turbulent diamagnetism,
reducing the magnitude of the magnetic field throughout the
experiment.  A previous paper showed that such currents must exist
\cite{Spence.PRL.2006}, but gave no information about their structure.
Here, direct measurement of the experiment's magnetic and velocity
fields is used to explicitly examine the structure of Ohm's law.

The Madison Dynamo Experiment is a one-meter-diameter sphere of
flowing liquid sodium~\cite{Nornberg.PP.2006}.  Sodium is chosen as
the working fluid for its high electrical conductivity.  An
axisymmetric mean velocity field is generated within the sodium by a
pair of counter-rotating impellers attached to shafts that enter the
sphere through each pole (with the shafts defining the axis of
symmetry).  The flowing sodium is very turbulent, with a kinetic
Reynolds number $Re\sim 10^7$.  The unconstrained geometry of the
experiment allows fluctuations in the velocity field to develop up to
the scale of the device.  The fluctuating velocity field generates a
fluctuating magnetic field by advecting the mean magnetic field of the
experiment.

The nature of Ohm's law is explored by measuring the
magnetic-field-dependent terms in Eq.~\ref{eqn:eqn1} to determine if
the induced magnetic field is due solely to the action of the mean
flow.  An approximately-uniform magnetic field is applied to the
flowing sodium using a pair of external magnetic field coils, and the
total magnetic field, applied plus induced, is measured (the applied
field is sufficiently weak that the velocity field is unaffected).
The collection of magnetic field data has been described
previously~\cite{Spence.PRL.2006}.  Since both the mean velocity and
applied magnetic fields are axisymmetric, the mean induced field is
also axisymmetric (all data presented in this Letter are
axisymmetric).  A spherical harmonic expansion of the induced internal
magnetic field, in the poloidal and toroidal directions, is fitted to
the most probable values of the measured magnetic field.  Since the
sphere is simply connected, a toroidal magnetic field cannot be
applied from outside the sphere; all measured toroidal magnetic field
is due to electrical currents flowing within the sodium.

\begin{figure*}
\includegraphics[height=3in]{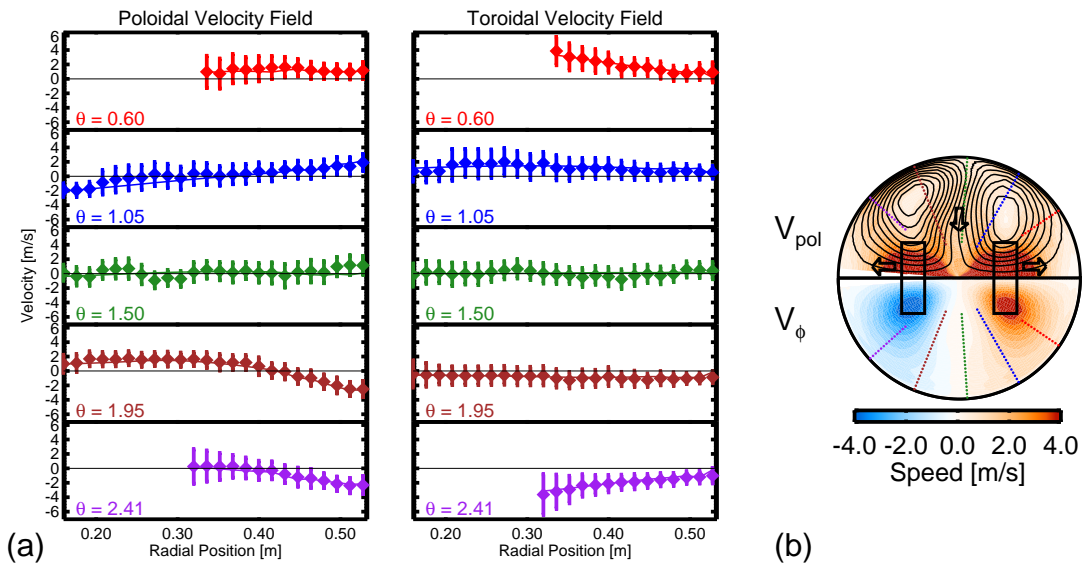}
\caption{Velocity fields measured by Laser Doppler Velocimetry in the
water model of the Madison Dynamo Experiment.  (a) Mean measured
velocity field as a function of radial position, for an impeller
rotation rate of 1000 RPM. (b) The reconstructed field.  The poloidal
flow rolls inward at the equator and outward at the poles.  Two
toroidal cells rotate in opposing directions in each hemisphere.  In
both panels the conventions are as in Fig.~\ref{fig:mag_fit}.}
\label{fig:vel_fit}
\end{figure*}

For an impeller rotation rate of 1000 revolutions per minute (RPM),
the mean measured induced magnetic field ranges from 1.2 times the
magnitude of the applied field in the poloidal direction, to 1.1 times
in the toroidal direction [Fig.~\ref{fig:mag_fit}(a)].  The field
reconstructed from the fit indicates that the external poloidal
magnetic field is dominated by a dipolar component that opposes the
externally-applied magnetic field [Fig.~\ref{fig:mag_fit}(b)].  The
reconstruction also demonstrates that the toroidal velocity field is
effective at generating toroidal magnetic field from the applied
poloidal magnetic field.

To distinguish between
$\left<\mathbf{B}\right>_{\left<\mathbf{V}\right> \times
\left<\mathbf{B}\right>}$, the magnetic field induced by the mean flow
($\nabla \times \left<\mathbf{B}\right>_{\left<\mathbf{V}\right>
\times \left<\mathbf{B}\right>} =
\mu_0\left<\mathbf{J}\right>_{\left<\mathbf{V}\right> \times
\left<\mathbf{B}\right>} \sim \mu_0\sigma \left<\mathbf{V}\right>
\times \left<\mathbf{B}\right>$), and
$\left<\mathbf{B}\right>_{\left<\mathbf{\tilde{V}} \times
\mathbf{\tilde{B}}\right>}$, the magnetic field due to fluctuations
($\nabla \times
\left<\mathbf{B}\right>_{\left<\mathbf{\tilde{V}} \times
\mathbf{\tilde{B}}\right>} \sim \mu_0\sigma \big<\mathbf{\tilde{V}}
\times \mathbf{\tilde{B}}\big>$), the
mean velocity field of the sodium must be known.  To this end an
identical-scale water model of the sodium experiment has been
constructed \cite{Forest.MHD.2002}.  At the correct temperatures water
and sodium have the same kinematic viscosity and similar densities.
As a result the two fluids are nearly hydrodynamically identical and
water can be used to model the flowing sodium.  Like the sodium
apparatus, the water model is a one-meter-diameter sphere in which
impellers generate an axisymmetric mean velocity field.  Stainless
steel tubes, identical to those containing the internal Hall probes in
the sodium experiment~\cite{Nornberg.PP.2006}, enter the flow at the
same seven locations as in the sodium experiment.  Unlike the sodium
apparatus, the water model is outfitted with five windows that allow a
Laser Doppler Velocimetry (LDV) system to directly measure the
poloidal and toroidal components of the water's velocity field.  A
spherical harmonic expansion of the velocity field is fitted to the
mean values of these measurements to determine the mean velocity field
of the flowing sodium.  For a impeller rotation rate of 1000 RPM, the
measured velocity field data yield a maximum poloidal speed of 2.5 m/s
and a maximum toroidal speed of 3.9 m/s [Fig.~\ref{fig:vel_fit}(a)].
As expected, the flow is counter-rotating in the toroidal direction
and the poloidal flow rolls inward at the equator and outward at the
poles [Fig.~\ref{fig:vel_fit}(b)].

\begin{figure*}
\includegraphics[height=3in]{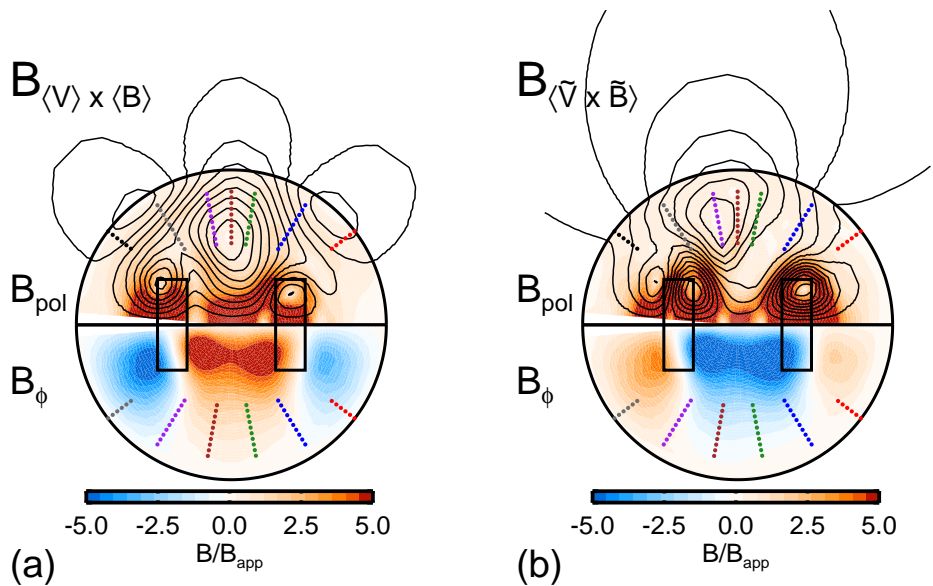}
\caption{Magnetic field due to the terms in Ohm's law.  (a) The
magnetic field, $\left<\mathbf{B}\right>_{\left<\mathbf{V}\right>
\times \left<\mathbf{B}\right>}$ calculated from the mean measured
velocity field interacting with the measured mean magnetic field (the
sum of the induced and the applied fields).  Note the lack of induced
external dipole moment.  (b) The mean magnetic field,
$\left<\mathbf{B}\right>_{\left<\mathbf{\tilde{V}} \times
\mathbf{\tilde{B}}\right>}$, due to the EMF associated with the
fluctuations.  The poloidal external field displays the measured
dipole component, in opposition to the applied field.  The toroidal
field is in opposition to the field induced by the mean flow of
sodium.}
\label{fig:emf}
\end{figure*}

Once the mean velocity field is known the magnetic field due to the
mean flow interacting with the mean magnetic field
[Fig.~\ref{fig:emf}(a)] and the magnetic field driven by the
fluctuations [Fig.~\ref{fig:emf}(b)] are calculated.  Like the fit to
the measured magnetic and velocity fields, this calculation is done in
a spherical harmonic basis, and involves calculating the electrical
potential of the experiment assuming
$\nabla\cdot\left<\mathbf{J}\right>=0$.  If no fluctuation-driven
currents were present ({\it i.e.~}the magnetic field is solely
explained by $\left<\mathbf{J}\right>_{\left<\mathbf{V}\right> \times
\left<\mathbf{B}\right>} \sim \sigma\left<\mathbf{V}\right> \times
\left<\mathbf{B}\right>$), then the measured mean magnetic field
[Fig.~\ref{fig:mag_fit}(b)] would be the same as the field due to the
mean velocity field [Fig.~\ref{fig:emf}(a)].  Inspection of these two
figures reveals that this is not the case.  The prominent external
dipole component of the measured poloidal magnetic field is completely
absent from the magnetic field due to the mean velocity field, since
the mean axisymmetric velocity field is incapable of producing it
\cite{Spence.PRL.2006}.  Also, the magnitude of the measured toroidal
magnetic field is significantly weaker than the toroidal field due to
the mean velocity field interacting with the measured magnetic field.
Clearly, the inductive action of the mean flow alone is insufficient
to explain the measured fields: fluctuations must be generating
significant magnetic field.
 
The magnetic field generated by the fluctuations has several prominent
features [Fig.~\ref{fig:emf}(b)].  First, there is a dipole component
that dominates the magnetic field outside the sphere, in the direction
opposite the applied field.  This is the source of the measured
external dipole moment.  Second, the toroidal field induced by the
fluctuations is in opposition to the field induced by the mean
velocity field interacting with the measured magnetic field.  Thus,
the field induced by the fluctuations is diamagnetic with respect to
the dominant poloidal and toroidal fields within the experiment.  The
effect is important to the overall magnetic field of the experiment,
as the fluctuation-driven field is a significant fraction of both the
field induced by the mean flow, and the applied field.  The strength
of the toroidal diamagnetic field near the poles is about 50\% of the
magnitude of the field induced by the mean flow, while the
fluctuation-driven dipole component is 20\% of the magnitude of the
applied field.  The fluctuation-driven poloidal field reduces the
total poloidal flux through the equatorial plane by 10\%.

Characterization of the fluctuations which lead to this turbulent EMF
requires direct measurement of the fluctuating components of
$\mathbf{V}$ and $\mathbf{B}$.  Since such measurements are not
currently available, it is not yet possible to determine whether the
observed diamagnetism is due to gradients in fluctuation levels,
helical turbulence, or some other effect.  Since the fluctuation
levels of the experiment are greatest near the impellers, and weakest
near the sphere's surface, flux expulsion due to gradients in
turbulence levels is a natural candidate to explain the observed
diamagnetism, though it is not the only possible explanation.

We note that the structure of the diamagnetic field is qualitatively
similar to the fluctuation-driven field predicted by recent numerical
simulations of the experiment~\cite{Bayliss.PRE.2006}.  These
simulations predict both the presence of the dipole moment and the
overall weakening of the measured toroidal field.  However, the
magnitude of the fluctuation-driven field presented here is a factor
of five larger than that predicted by the simulations.  The reason for
this discrepancy is not currently known, though it may be related to
differences in fluid forcing or the simulation's large value of
magnetic Prandtl number.

In summary, mean magnetic and velocity field measurements have been
used to determine the structure of fluctuation-driven currents in the
Madison Dynamo Experiment.  These currents lead to a magnetic field in
opposition to the experiment's dominant magnetic field.  This is the
first observation of turbulent diamagnetism in a laboratory setting.
Analysis of the nature of the fluctuations which leads to this
turbulent EMF is ongoing, and future work will focus on further
understanding the localized regions of current generation.  Given that
the observed effect is diamagnetic, it indicates that such fields
could be a hindrance to the development of magnetically self-exciting
geophysical and astrophysical systems, as well as simply-connected
liquid-metal dynamo experiments.

E.J.S. thanks A. Jackson and C. Finlay for reviewing the manuscript.
This work is funded by the US Department of Energy, the National
Science Foundation, and David and Lucille Packard Foundation.

\end{document}